%% file: main.tex
\documentclass[twocolumn,12pt]{aastex6}
\usepackage{hyperref}
\usepackage{amsmath}
\usepackage{amssymb}
\usepackage{graphicx}
\usepackage{natbib}
\usepackage{multirow}
\usepackage{rotating}
\usepackage{color}
\usepackage{float}
\usepackage{array}

\DeclareRobustCommand{\ion}[2]{\textup{#1\,\textsc{\lowercase{#2}}}}
\newcommand\kms{\ensuremath{\mbox{km}\,\mbox{s}^{-1}}}
\newcommand\Teff{\ensuremath{T_\mathrm{eff}}}
\newcommand\logg{\ensuremath{\log g}}

\newcommand\vt{\ensuremath{\xi_{t}}}

\newcommand\fei{\ion{Fe}{i}}
\newcommand\feii{\ion{Fe}{ii}}
\newcommand\feiii{\ion{Fe}{iii}}

\begin{document}
\shorttitle{NLTE stellar parameters of UMP stars}
\title{Ultra-Metal poor stars: Spectroscopic determination of  stellar atmospheric parameters using iron non-LTE line abundances}

\author{Rana Ezzeddine\altaffilmark{1,2},
Anna Frebel\altaffilmark{2,1} and Bertrand Plez\altaffilmark{3}}

\altaffiltext{1}{Joint Institute for Nuclear Astrophysics, Center for the Evolution of the Elements, East Lansing, MI 48824, USA}
\altaffiltext{2}{Department of Physics and Kavli Institute for Astrophysics and Space Research, Massachusetts Institute of Technology, Cambridge, MA 02139, USA}
\altaffiltext{3}{Laboratoire Univers et Particules de Montpellier, Universit\'e
de Montpellier, CNRS, UMR 5299, Montpellier, France}

\begin{abstract}
We present new ultra-metal-poor (UMP) stars parameters with [Fe/H]\,$<$\,$-$4.0 based on line-by-line non-local thermodynamic equilibrium (NLTE) abundances using an up-to-date iron model atom with a new recipe for non-elastic hydrogen collision rates.  We study the departures from LTE in their atmospheric parameter and show that they can grow up to $\sim 1.00$\,dex in [Fe/H], $\sim 150$\,K in $T_{\mathrm{eff}}$ and $\sim 0.5$\,dex in log\,$g$ toward the lowest metallicities. Accurate NLTE atmospheric stellar parameters, in particular [Fe/H] being significantly higher, are the first step to eventually providing full NLTE abundance patterns that can be compared with Population\,III supernova nucleosynthesis yields to derive properties of the first stars. Overall, this maximizes the potential of these likely second-generation stars to investigate the early universe and how the chemical elements were formed. 

\end{abstract}


\section{Introduction} \label{sec:intro}
Ancient ultra-metal-poor (UMP) stars (with $\mbox{[Fe/H]}<-4.0$, e.g. \citealt{Beers2005}) are rare relics of the early Universe. They provide unique insights into the first nucleosynthesis events and the first (Population\,III; Pop\,III hereafter) stars \citep{Klessen2012,Bromm2013}, the earliest phases of chemical enrichment \citep{Frebel2015}, as well as the formation of the first low-mass stars \citep{frebel2007,chiaki2012,ji2015}. For example, detailed comparisons of supernova nucleosynthesis yields with stellar abundances have shown that Pop\,III stars were likely massive (20-60\,M$_{\odot}$; \citealt{keller2014,tominaga2014, Placco2015}), in agreement with theoretical expectations \citep{bromm2002,abel2001}. As more stars and more detailed yield calculations become available, the nature and shape of the initial mass function of Pop\,III stars can ultimately be reconstructed that way.

Key ingredients are as accurate and precise as possible stellar abundances of many elements. For example, \citet{Placco2015} investigated how abundance availability and precision affected the results of fitting the abundance patterns with nucleosynthetic yields to derive Pop\,III stellar masses. They found that the exclusion of nitrogen from the abundance pattern had a significant impact on the final derived Pop\,III progenitor mass. To obtain high quality chemical abundances, a necessary prerequisite is accurate and precise stellar atmospheric parameters, i.e., effective temperature \Teff, surface gravity \logg, iron abundance [Fe/H], and microturbulence \vt\ that characterize the star. Stellar parameters can be determined in different ways, but in addition to the metallicity, in most cases as least the microturbulence and/or the surface gravity are determined spectroscopically by demanding no abundance trend with reduced equivalent width and invoking ionization equilibrium of neutral and singly ionized iron lines, \fei\ and \feii, respectively. Effective temperature can also be determined spectroscopically, by invoking the ``excitation balance'', i.e. no abundance trend with excitation potential of \fei\ lines.

For most abundance analyses, one-dimensional stellar model atmospheres are used, together with radiative transfer codes assuming local thermodynamic equilibrium (LTE). This method, however, is affected by unaccounted departures from LTE that can introduce significant systematic uncertainties since line formation and populations of non-dominant species (in this case \fei) can potentially deviate from Saha-Boltzmann equilibrium assumed in LTE (e.g. \citealt{rutten2003}). 

Deviations from LTE have been shown to increase toward lower metallicities and for extended atmospheres (giants and super-giants) \citep{mashonkina2011,bergemann2012,lind2012,Bergemann2015,mashonkina2016}. The decreasing number of electrons donated by metals leads to decreased collision rates. In cool, late-type stars (4000\,K $< \Teff < 6500$\,K), lines arising from minority species are most affected by these deviations. 
To account for such departures, especially in cool UMP stars, it is necessary to investigate the formation of iron lines in non-local thermodynamic equilibrium (non-LTE, hereafter NLTE). For such an investigation, atomic data and other inputs are required for each element. Previous iron NLTE studies have reported abundance changes ($\sim$ 0.1 dex) simply due to uncertainties in the input atomic data used in their model atoms \citep{bergemann2012, mashonkina2011, collet2005}. Particularly problematic in this context are uncertainties arising from estimating the rate of inelastic collisions with neutral hydrogen atoms. They are usually obtained from the classical Drawin approximation \citep{drawin1968,drawin1969a,drawin1969b,lambert1993} because  full quantum calculations are lacking. However, the approximation is known to overestimate the collision rates by several orders of magnitude \citep{barklem2010,lind2011,osorio2015}. Several attempts to calibrate these rates were made by applying a global multiplicative fudge factor (denoted $\mbox{S}_{\mathrm{H}}$) to all the rates, calibrated against different benchmark stars. However, unlike what is expected for an intrinsic atomic property, different $\mbox{S}_{\mathrm{H}}$ were obtained by different studies which were found to be star and model atom dependent (e.g. \citealt{thevenin1999,korn2003,mashonkina2016,bergemann2012,Ezzeddine2016}). 

To obtain more accurate collision rates, a new semi-empirical recipe was proposed by \citet{Ezzeddine2016b} who developed a quantum fitting method (QFM) to estimate the hydrogen collision rates involved in iron line formation and other elements in the lack of available published quantum rates. Transition energies dependent recipes for charge transfer and excitation rates were introduced to determine more reliable Fe abundances. The QFM has already been successfully applied to line formation calculations in solar-type FGK giants and metal-poor stars, such as the Sun, HD140283 and Arcturus. In this paper, we apply the QFM and our Fe model atom to 20 known UMP stars with $\mbox{[Fe/H]}\lesssim-4.0$. We use high-resolution spectra available in the literature to spectroscopically determine their stellar parameters using NLTE calculations of available Fe lines. This is the first step toward eventually determining their full NLTE abundance patterns.

The paper is structured as follows: In Section~\ref{sec:Models and observational data}, we describe the input data used in the NLTE calculations, the iron model atom, and input atomic linelists and model atmospheres. In Section~\ref{sect:method}, we introduce the method used. Both our LTE and NLTE iron abundances are presented in Section~\ref{results}, as well as the light element enhancement effects on the final results. In Section~\ref{NLTE_corr}, we report the NLTE corrections for the LTE iron line abundances, as well as implications for the stellar parameters of the UMP stars. Finally, conclusions are presented in Section~\ref{conc}.

\section{Input data for NLTE line formation calculations} \label{sec:Models and observational data}

\subsection{Iron model atom}\label{iron_model}
The iron model atom used in this work was built from all available energy levels for \fei\ and \feii\ (846 \fei\ and 1027 \feii\ levels and the \feiii\ continuum) from the \texttt{NIST}\footnote{https://www.nist.gov/pml/atomic-spectra-database} database which were then collapsed into superlevels, disregarding fine-structure splitting for all levels except those of \fei\ and \feii\ ground levels. The model also includes the predicted high lying \fei\ levels from \citet{petkur2015} that correspond to UV and IR transitions, and establish important collisional couplings with the ground \feii\ level. All levels in the atom are coupled via an extensive linelist of radiative bound-bound (extracted from \texttt{VALD3}\footnote{http://vald.astro.uu.se/}  database) and photoionization transitions (extracted from the \texttt{NORAD}\footnote{http://www.astronomy.ohio-state.edu/$\sim$csur/NORAD/norad.html} database). A detailed description of the iron model atom can be found in \citet{Ezzeddine2016b}.

Additionally, all levels in the atom are coupled via electron and hydrogen collisions. These collisions have been shown to have important effects on the final NLTE abundances (an extensive study showing this for Mg is presented in \citealp{osorio2015}). While quantum atomic data for the hydrogen collision cross-sections have been computed for some light atoms (Li, Be, Na, Mg, Al, Si and Ca; \citealp{belyaev2003}, \citealp{yakovleva2016}, \citealp{barklem2010}, \citealp{belyaev2012}, \citealp{belyaev2013}, \citealp{belyaev2014} and \citealp{belyaev2016} respectively), data for larger atoms remain scarce and more difficult to compute. 

In ultra metal-poor stellar atmospheres, the hydrogen to electron density ratios can reach up to $\sim 10^{5}$ due to the scarcity of free electron donors (i.e., metals). This enhances the role that inelastic hydrogen collision rates can play for a NLTE abundance determination. Recently, it has been shown that the charge-transfer (i.e., ion-pair production\footnote{During an atomic collision with hydrogen, i.e., A+H, the valence electron associated with the atom A has a certain probability to tunnel into the H atom, resulting in a predominantly ionic charge distribution or an ion-pair production A$^{+}$+H$^{-}$.}) processes can dominate over excitation processes \citep{osorio2015,osorio2016,Ezzeddine2016b}. Cross-section calculations for Li+H \citep{barklem2003} and Na+H \citep{barklem2010}, for example, have shown that the largest cross sections for excitation are small compared with those for ion-pair production from certain energy states. Ion-pair production cross-sections calculations for Fe have not been published yet and were thus excluded in most previous NLTE iron studies for UMP stars.

  A new semi-empirical quantum fitting method (QFM) to estimate the hydrogen collision rates, including the ion-pair production process, was developed by \citet{Ezzeddine2016b}.  This method is based on a general fitting recipe deduced from the quantum collision rates of several elements (Be, Na, Mg, Al, Si and Ca) and then applied to Fe. Tested on 24 \textit{Gaia} benchmark stars \citep{jofre2014a,heiter2015} with different stellar parameters, it was shown to improve the \fei\ and \feii\ ionization balance and decrease the obtained abundance scatter, especially for the more metal-poor stars in the sample. This motivates the present work with UMP stars which are expected to experience significant NLTE abundance effects.

\subsection{Linelists and equivalent widths}\label{linelists}
For our Fe abundance determination, we use absorption lines and equivalent widths measurements for each UMP star from relevant references in the literature. Details are presented in Section~{\ref{ump_stell_param}} below. $\mbox{Log}\,gf$ values for \fei\ and \feii\ lines from Gaia-ESO ``golden'' linelist \texttt{v4} \citep{heiter2015b} were used. The linelist used for each star can be found in Table~{\ref{Tab:abund}}.

\subsection{Model atmospheres}\label{mod_atmos}
We employ 1D, plane-parallel MARCS atmospheric models \citep{gustafsson1975,gustafsson2008,plez2008} which were interpolated\footnote{The interpolation routine \texttt{interpol\_modeles.f} from Thomas Masseron available on http://marcs.astro.uu.se/software.php was used.} to the corresponding input stellar parameters for each star listed in Table~{\ref{tab:ump_list_param}}. We use models with a metallicity of $-5.0$ dex for all stars with $\mbox{[Fe/H]}\leq-5.0$. Standard $\alpha$-element enhancement of $\mbox{[$\alpha$/Fe]}=0.4$ was adopted for all UMP stars.
Blanketing effects were taken into account by including background line opacity tables (excluding Fe) as a function of $\mbox{[Fe/H]}$ and $\vt$\ (B. Plez priv. communication). Throughout, we adopt a reference solar iron abundance of  $\log \epsilon$(Fe)$_{\odot}$ = 7.50 from \citet{asplund2009}.

\section{Method}\label{sect:method}
We use the input data introduced in Section~{\ref{sec:Models and observational data}} (including the Fe model atom, atomic linelists and 1D MARCS atmospheric models) to determine \fei\ and \feii\ line-by-line NLTE abundances. We also report the LTE Fe abundance for each UMP star.

The NLTE radiative transfer code \texttt{MULTI2.3} \citep{Carlsson1986,carlsson1992} was used to compute NLTE line profiles using the Accelerated Lambda Iteration (ALI) approximation method \citep{scharmer1981}. It computes the level populations by solving the statistical equilibrium and radiative transfer equations simultaneously with no feedback from the element in question back into the atmosphere (i.e. considered as a trace element). For each line, the NLTE equivalent width $EW_{\mathrm{NLTE}}$ was computed using a Voigt profile function with a maximum of 80 frequency points. We also compute the LTE equivalent width $EW_{\mathrm{LTE}}$  from the departure coefficient of each line $i$ such as $b^{i}=EW^{i}_{\mathrm{NLTE}}/EW^{i}_{\mathrm{LTE}}$ \citep{wijbenga1972}. A curve-of-growth (COG) method is then used to determine the line abundances that correspond to the observed $EW_{\mathrm{obs}}$. All lines used in the abundance analysis lie on the linear part of the COG ($\log (EW_{\mathrm{obs}}/\lambda) < -4.8$). 

\subsection{Stellar atmospheric parameters}\label{ump_stell_param}
We determined NLTE spectroscopic atmospheric parameters for the UMP stars, using NLTE calculations of the abundances of individual \fei\ and \feii\ lines and upper limits following the method outlined in Section~{\ref{sect:method}}. To guide our calculations, we make use of the fact that stellar parameters for all stars have previously been determined under the assumption of LTE, either fully spectroscopically (\Teff, \logg, \vt\ and [Fe/H]) or partially (either \Teff\ or \logg, \vt\ and [Fe/H]). We first computed a small grid of NLTE $EW$ at stellar parameters centered around the LTE (or photometric) stellar parameters from the literature (see Section~\ref{results}). We then compared the corresponding grid of computed NLTE equivalent widths, $EW_{\mathrm{NLTE}}$, with the measured observed ones $EW_{\mathrm{obs.}}$.  This was done for all stars for which $EW_{\mathrm{obs.}}$  of at least 5 iron lines could be measured. A first approximation of the initial stellar parameters for each star (in terms of \Teff, \logg, [Fe/H]) was obtained using non-linear  $\chi^{2}$ fitting of the computed $EW_{\mathrm{NLTE}}$ to the observed $EW_{\mathrm{obs.}}$ in a \Teff-\logg-[Fe/H] parameter space. This procedure (using a Levenberg-Marquardt algorithm) takes into account that all stellar parameters depend on each other.

In a second step, the excitation balance of \fei\ lines abundances as a function of excitation potential of the lower level, $\chi$, of each line, as well as the ionization equilibrium of \fei\ and \feii\ abundances (when available) were inspected. In the case of an abundance trend with $\chi$ or a mismatch between \fei\ and \feii\ abundances, stellar parameters were adjusted accordingly. In the process, we derived the microturbulent velocity $\xi_{t,\mathrm{NLTE}}$ by removing any \fei\ line abundance trend with reduced equivalent widths ($\log(EW_{\mathrm{obs}}/\lambda)$).

\input{Stellar_params.tex}

\section{Results}\label{results}
To proceed with our analysis, we divided our sample of 20 UMP stars into three metallicity subgroups following the classification in \citet{Placco2015} as shown in Table~\ref{tab:ump_list_param}: Hyper-metal-poor (HMP) stars with $\mbox{[Fe/H]}<-5.0$, stars with $-5.0<\mbox{[Fe/H]}<-4.5$ and stars with $-4.5<\mbox{[Fe/H]}<-4.0$. We exclude the carbon rich star G~77$-$61 from our UMP sample due to the complexity of its spectrum showing very strong CH, CN, and C2 bands around all iron line regions. Its analysis would require the inclusion of CN and C2 lines in the continuum and line background opacities in the NLTE analysis code which is beyond the scope of the present work. Below we present a brief description of  all stars in each subgroup and our stellar parameters in NLTE. We also discuss differences to LTE stellar parameters. All respective parameters are listed in Table~{\ref{tab:ump_list_param}}. Our final NLTE Fe abundances and associated NLTE corrections are further analyzed in Sections~\ref{results} and \ref{NLTE_corr} respectively.

\subsection{HMP stars with \normalfont[Fe/H] $<-5$}
This group includes five stars, SMSS~J0313$-$6708, HE~1327$-$2326, HE~0107$-$5240, SDSS~J1035$+$0641 and SDSS~J1313$-$0019, with the lowest iron abundance of $\mbox{[Fe/H]} < -5.0$, as determined from their LTE analysis. 

\textbf{SMSS~J0313$-$6708} is a warm red giant with the lowest known iron abundance \citep{keller2014}. Only an upper limit of the Fe abundance could be derived because no Fe lines were detected in the spectrum.
The authors determined $T_{\mathrm{eff}} = 5125 \pm 100$\,K and $\log g = 2.3\pm 0.2$ from spectro-photometry, consistent with results from stellar hydrogen line profiles and the derived lithium abundance. Using an equivalent width upper limit of the strongest Fe\,I lines (at 3859.91\,{\AA}), \citet{keller2014} determined an upper iron abundance limit of $\mbox{[Fe/H](LTE)}<-7.30$. Also employing a $\langle \mbox{3D} \rangle$, NLTE correction for this line from \citet{lind2012}, led to $\mbox{[Fe/H](NLTE)}<-7.10$. \citet{bessell2015} redetermined the upper limit in NLTE to $\mbox{[Fe/H](NLTE)} < -7.52 \pm 1\sigma$ using a $\langle\mbox{3D}\rangle$ model atmosphere and a spectrum with higher signal-to-noise. A microturbulent velocity of 2.0\,{\kms} was adopted for the star in both studies.
More recently, \citet{nordlander2017} performed a full 3D, NLTE analysis of this star using up to date atomic and hydrogen collisional data independent of classical approximations and free parameters. This led to higher iron abundances than \citet{bessell2015}, of $\mbox{[Fe/H](1D,NLTE)}<-6.73$ and $\mbox{[Fe/H](3D,NLTE)}<-6.53$ by fitting a stacked spectra in the vicinity of unblended \fei\ lines at 3440.6\,{\AA}, 3581.2\,{\AA}, 3719.9\,{\AA}, 3737.1\,{\AA}, 3820.4\,{\AA} and 3859.9\,{\AA} respectively. This discrepancy with the \citet{bessell2015} value was explained being due to differences in atomic data and the use of full 3D model as compared to an averaged $\langle \mbox{3D} \rangle$ model.

In this work, we adopt an upper limit for $EW_{\mathrm{obs}} < 1.0$\,m{\AA} for the strongest \fei\ lines at 3608.859\,{\AA} and 3859.911\,{\AA} respectively. 
We find $\mbox{[Fe/H](LTE)}<-7.79$ using only the resonance line at 3859.911\,{\AA}. We also compute $\mbox{[Fe/H](LTE)}<-7.24$ from the non-resonance line at 3608.859\,{\AA}. Our $3859.911$\,{\AA} result agrees with that of \citet{bessell2015}, $\mbox{[Fe/H](LTE)}<-7.80$, who used the same line. Our 3608.859\,{\AA} LTE value agrees with that of \citet{nordlander2017} within 0.1\,dex (they report $\mbox{[Fe/H](1D,LTE)}<-7.34$).
In NLTE, we determine an upper limit of $\mbox{[Fe/H](NLTE)}<-6.52$ from the 3608.859\,{\AA} line, and $\mbox{[Fe/H](NLTE)}<-6.72$ from the resonance 3859.911\,{\AA}.
The abundance obtained from the resonance line is 0.2\,dex lower than the 3608.859\,{\AA} line. We thus adopt the upper limit of $\mbox{[Fe/H](NLTE)}<-6.72$ from the resonant line at 3859.911\,{\AA} as our final result. This leads to an iron abundance that is 0.80\,dex higher than the $\langle$3D$\rangle$ NLTE result determined by \citet{bessell2015} (who reported $\mbox{[Fe/H]}<-7.52$). This result is in perfect agreement with the 1D, NLTE result of \citet{nordlander2017} (who reported $\mbox{[Fe/H]}<-6.73$). Our 1D, NLTE result is, however,  0.19\,dex lower than their 3D, NLTE value. Overall, \citet{nordlander2017} report agreement between their 1D and 3D NLTE results which adds confidence to our result and the use of our NLTE method. As no Fe lines could be detected, we did not compute any NLTE stellar parameters but instead adopt the temperature, gravity and microturbulent velocity from \citet{bessell2015}.

\textbf{HE~1327$-$2326} is a relatively unevolved star located on either the main-sequence or the subgiant branch \citep{frebel2005, aoki2006,frebel2008}. \citet{frebel2005} used color-effective temperature relations from \citet{alonso1996}, to determine  $T_{\mathrm{eff}}= 6180\pm80$\,K from broad-band $UBVRI$ photometry. They used the proper motion to set limits on the distance, and from a 12~Gyr isochrone with $\mbox{[Fe/H]} = -3.5$, two solutions, $\log g =3.7$ and 4.5, were obtained. \citet{korn2008} favored a subgiant scenario after carrying out a NLTE \ion{Ca}{i}/\ion{Ca}{ii} ionization equilibrium analysis. The iron abundance of HE~1327$-$2326 was determined using 10 \fei\ lines from \citet{frebel2008} as no \feii\ lines could be detected. For the subgiant case, \citet{frebel2008} derived $\mbox{[Fe/H](1D,LTE)} = -5.71 \pm 0.2$ and $\mbox{[Fe/H](3D,LTE)}= -6.01 \pm 0.2$. A nominal NLTE correction of 0.2\,dex (without any tailored calculation) was adopted in \citet{frebel2005} following \citet{asplund2005}. A microturbulent velocity of $\vt = 1.7$\,{\kms} was adopted throughout.

We use 10 \fei\ lines from \citet{frebel2008} for our analysis of HE~1327$-$2326, and additionally a strong \feii\ line at 5018.45\,{\AA} for which we use an upper limit of $EW_{\mathrm{obs}}<0.8$\,m{\AA}. 
Applying our stellar parameters fitting method described in Section~{\ref{ump_stell_param}}, we find a best fit at \logg=3.7, \Teff=6130\,K but adopt $\vt = 1.7$\,\kms\ as in \citet{frebel2008}, given the paucity of lines. These values satisfy both the excitation and ionization equilibrium (to the extent the upper limit allows). We thus also favor the subgiant scenario, in agreement with \citet{korn2008}. Our \Teff\ result agrees well with that of \citet{frebel2005}, however, detecting and measuring any \feii\ lines in this star would provide a better constraint on \logg. Using our derived stellar parameters, we determine two sets of abundances for each \logg\ scenario. As such, we derive iron abundances of 
$\mbox{[Fe/H](LTE)}=-5.82$ and $\mbox{[Fe/H](NLTE)}=-5.16$ for the subgiant case, and $\mbox{[Fe/H](LTE)}=-5.76$ and $\mbox{[Fe/H](NLTE)}=-5.22$ for the dwarf case. Our subgiant LTE Fe abundance agrees with \citet{frebel2008} within acceptable 0.11\,dex.

\textbf{HE~0107$-$5240} is a red giant star \citep{christlieb2002}. \citet{christlieb2004} derived $T_{\mathrm{eff}} = 5100 \pm 150$\,K following $(b-y)$ - $T_{\mathrm{eff}}$ relations by \citet{alonso1999,alonso2001}. They used different methods including relative strengths of Balmer line wings and evolutionary tracks to constraint the surface gravity. $\logg = 2.2 \pm 0.3$\,dex was eventually adopted. $EW$ of 25 \fei\ lines were measured and one upper limit of $EW<10$\,m{\AA} for the \feii\ line at 5018.440\,{\AA}. $\vt = 2.2 \pm 0.5$\,{\kms} was determined by forcing the abundances of \fei\ lines to have no trend with line strengths. This led to iron abundances of $\mbox{[Fe/H](LTE)} =-5.44 \pm 0.2$. Adopting a nominal NLTE correction of 0.11\,dex, they report $\mbox{[Fe/H](NLTE)}=-5.35 \pm 0.2$ (without carrying any detailed NLTE calculation).

 Using our $EW$ fitting method, we determine atmospheric parameters of $T_{\mathrm{eff}} =5050$\,K, $\logg = 2.3$ and $\vt = 2.2$\,\kms, in good agreement with those presented in \citet{christlieb2004}.
Our LTE abundance of $\mbox{[Fe/H](LTE)} = -5.47$ is in very good agreement with that of \citet{christlieb2004}. We then determine 
 $\mbox{[\fei/H](NLTE)} = -4.72$ and $\mbox{[\feii/H](NLTE)} < -4.71$ from the same lines as in \citet{christlieb2004}. We note that the upper limit for the \feii\ line ($\lambda$5018.44\,{\AA}) is already at the level of the \fei\ abundance. Should the true \feii\ abundance be significantly lower, the surface gravity of the star would need to be significantly increased.
 
\textbf{SDSS~J1035$+$0641} is a warm dwarf star \citep{bonifacio2015}. No metal lines were found in its spectrum except for the \ion{Ca}{ii}\,K line and the G-band. \citet{bonifacio2015} derived $T_{\mathrm{eff}}=6260$\,K from a $(g - z)$ calibration, found to be consistent with what was determined from the H${\alpha}$ line wings. Using a 12 Gyr isochrone, two possible values for \logg\ of 4.0 and 4.4 were found. An upper iron abundance limit of $\mbox{[Fe/H](LTE)}<-5.59$ was set from synthesizing the wavelength region of 3820-3860\,{\AA} where the three strongest \fei\ lines are found. Following \citet{caffau2013}, they assumed $\vt= 1.5$\,\kms\, for the microturbulent velocity due to the lack of any Fe lines.

As no iron lines were detected in this star, we could not derive NLTE stellar parameters with our spectroscopic fitting method. 
Using the \Teff\, and both \logg\ values, and the $1\sigma$ upper limit on the equivalent width for the \fei\ line at 3820.425\,{\AA} ($EW_{\mathrm{obs}}<5.7$\,m{\AA}) from \citet{bonifacio2015}, we determine identical \fei\ upper limit abundances in LTE and NLTE for both cases of $\log g$ of $\mbox{[\fei/H](LTE)} < -5.72$ and $\mbox{[\fei/H](NLTE)} < -5.18$. The LTE upper limit agrees well with that of \citet{bonifacio2015} within 0.13\,dex. 
 
\textbf{SDSS~J1313$-$0019} is a star at the base of the red giant branch \citep{frebel2015b}. Its effective temperature $T_{\mathrm{eff}}= 5170 \pm 150$\,K was determined spectroscopically (LTE) using excitation balance and applying a temperature correction following \citet{frebel2013}. As no \feii\ lines could be detected in the spectrum, a surface gravity of $\logg = 2.6 \pm 0.5$ was obtained using a 12 Gyr isochrone at $\mbox{[Fe/H]}=-3.0$. Iteratively, a microturbulent velocity of $\vt = 1.8\pm0.3$\,{\kms} and iron abundance of $\mbox{[Fe/H]}=-5.0\pm0.1$ were determined.

We determine atmospheric parameters of $T_{\mathrm{eff}} =5100$\,K and $\vt = 1.8$\,{\kms}. As no \feii\ lines were detected, we could not derive \logg\ via ionization equilibrium, however with our fitting method we obtain a best fit at $\logg=2.7$. This value could be further investigated with the detection and measurement of \feii\ lines. We then used 36 \fei\ lines from \citet{frebel2015b} to determine iron abundances of $\mbox{[Fe/H](LTE)}=-5.02\pm0.09$ and $\mbox{[Fe/H](NLTE)}=-4.41\pm0.08$.

\subsection{UMP stars with $-5.0 <$ \normalfont{[Fe/H]} $< -4.5$}
These include SDSS~J1029+1729, SDSS~J1742+2531, HE~0557$-$4840 and HE~0233$-$0343. 

\textbf{SDSS~J1029+1729} is a Turn-Off (TO) star first analyzed by \citet{caffau2011,caffau2012}. They derived $T_{\mathrm{eff}}=5811\pm150$\,K from $(g - z)$ color relations from \citet{ludwig2008} and $\logg=4.0\pm0.5$ from \ion{Ca}{i}/\ion{Ca}{ii} ionization equilibrium (due to the lack of any \feii\ lines). They obtained $\vt = 1.5$\,{\kms} following the relation of \citet{edvardsson1993}. Iron abundances were derived as $\mbox{[Fe/H](1D,LTE)}=-4.71\pm0.13$ in LTE and $\mbox{[Fe/H](NLTE+3D corr.)}=-4.87\pm0.10$, as obtained from adding 3D corrections to a NLTE analysis. 

Only 3 \fei\ lines could be measured by \citet{caffau2012}. Due to the scarcity of lines, we did not derive any NLTE stellar parameters for this star. Using stellar parameters and the three \fei\ $EW$ measurements from \citet{caffau2012}, we derive $\mbox{[Fe/H](LTE)}=-4.63\pm0.13$ and $\mbox{[Fe/H](NLTE)}=-4.23\pm0.14$. Our LTE value is in agreement with that from \citet{caffau2012} within error bars. 

\textbf{SDSS~J1742+2531} is a warm TO star for which \citet{bonifacio2015} derived $T_{\mathrm{eff}} = 6345$\,K, from $(g-z)_{0}$ colors, and in agreement with that derived from H${\alpha}$ line wings by the authors. Two values for \logg\ were determined using a 12 Gyr, $Z=2\times10^{-4}$ isochrone: 4.0 for an evolved past turn-off (TO) star case and 4.3 for an un-evolved case.  $\logg=4.0$ was eventually adopted, arguing accordance with a metal-poor TO star at this temperature. Using three measured \fei\ lines, they derive $\mbox{[Fe/H](LTE)}=-4.78\pm0.08$.

We use the same lines to determine $\mbox{[Fe/H](LTE)}=-4.82\pm0.07$ and $\mbox{[Fe/H](NLTE)}=-4.34\pm0.03$. Due to scarcity of Fe lines, we refrain from determining other NLTE stellar parameters for this star, and adopt the temperature, gravity and $\vt$ from \citet{bonifacio2015}.

\textbf{HE~0557$-$4840} is an evolved red giant star  with  $T_{\mathrm{eff}}=4900$\,K, as determined from fitting Balmer lines \citep{norris2007}.  $\logg=2.2$ was determined from the \fei/\feii\ and \ion{Ca}{i}/\ion{Ca}{ii} ionization equilibrium, and subsequently $\vt=1.8$\,{\kms} was obtained. From these stellar parameters they derived $\mbox{[Fe/H](LTE)}=-4.80\pm0.2$.

Using 59 \fei\ and 1 \feii\ lines from \citet{norris2007}, we determine $T_{\mathrm{eff}} =4800$\,K, $\logg=2.4$ and $\vt=1.8$\,{\kms}. Subsequently, we derive $\mbox{[Fe/H](LTE)}=-4.86$ and $\mbox{[Fe/H](NLTE)}=-4.48$, where our LTE value agrees with that of \citet{norris2013}. 
                                          
\textbf{HE~0233$-$0343} is a warm subgiant, and one of three UMP stars first analyzed by \citet{hansen2014}.  $T_{\mathrm{eff}}=6100\pm100$\,K was derived by fitting spectro-photometric observations with 1D, LTE MARCS model atmospheric fluxes and $\logg=3.4\pm0.3$ by employing $\alpha$-element enhanced, 10 Gyr isochrone at $\mbox{[Fe/H]}=-4.7$. No \feii\ lines were detected in the spectrum. $\mbox{[Fe/H](LTE)}=-4.7\pm0.2$ and $\vt=2.0\pm0.3$\,{\kms} were derived from 11 \fei\ lines.

Using $EW$ measurements of \fei\ lines provided by T. Hansen (priv. communication), we determine in NLTE $T_{\mathrm{eff}} =6020$\,K, $\logg=3.4$ and $\vt=2.0$\,{\kms} using the $EW$ fitting method, and derive $\mbox{[Fe/H](LTE)}=-4.44\pm0.08$ and $\mbox{[Fe/H](NLTE)}=-3.99\pm0.06$.

\subsection{UMP stars with $-4.5<$ \normalfont{[Fe/H]} $<-4.0$}
These include CS~22949$-$037, SDSS~J1204+1201, CD$-$38 245, HE~1310$-$0536, HE~2239$-$5019, SDSS~J0140+2344, HE~0057$-$5959, HE~1424$-$0241, CS~30336$-$049, HE~2139$-$5432 and SDSS~J2209$-$0028.

\textbf{CS 22949$-$037} is a well studied red giant star \citep{beers1992,depagne2002,roederer2014}. \citet{beers1992} determined $\mbox{[Fe/H](LTE)}=-3.80$ from medium resolution spectrum, using the strength of the \ion{Ca}{ii}\,K line. This value was then re-determined by many authors, e.g. \citet{depagne2002} who derived $\mbox{[Fe/H](LTE)}=-3.94$, and \citet{roederer2014} who reported $\mbox{[Fe/H](LTE)}=-4.21$.
\citet{depagne2002} estimated $T_{\mathrm{eff}}=4900\pm125$\,K from giant stars color-calibrations relations from \citet{alonso1999}. $\logg=1.5$ was derived using \fei/\feii\ and \ion{Ti}{i}/\ion{Ti}{ii} ionization equilibrium. Finally, $\vt=1.8$\,{\kms} was deduced by minimizing the Fe abundance trend as a function of $\log(EW_{\mathrm{obs}}/\lambda)$.

We determine $T_{\mathrm{eff}} =4800$\,K, \logg=1.9 and $\vt=1.9$\,{\kms}, and subsequently derive $\mbox{[Fe/H](LTE)}=-3.99\pm0.16$ and $\mbox{[Fe/H](NLTE)}=-3.48\pm0.13$ from 65 \fei\ and 5 \feii\ lines from \citet{depagne2002}. Our LTE result agrees very well with that of \citet{depagne2002}.

\textbf{SDSS J1204+1201} is an evolved subgiant star \citep{Placco2015}. The authors determined $T_{\mathrm{eff}}=5467\pm100$\,K from excitation balance of \fei\ lines, $\logg=3.2\pm0.20$ by employing a 12 Gyr isochrone at $\mbox{[Fe/H]}=-3.5$ and $\vt=1.5\pm0.2$\,{\kms} from \fei\ lines balance with reduced $EW_{\mathrm{obs}}$. From the above parameters, they derive $\mbox{[Fe/H](LTE)}= -4.34\pm0.05$.

We employ our $EW$ fitting technique to determine $T_{\mathrm{eff}} =5350$\,K, $\logg=3.3$ and $\vt=1.5$\,{\kms}. $\mbox{[Fe/H](LTE)}=-4.39\pm0.12$ and $\mbox{[Fe/H](NLTE)}=-3.91\pm0.11$ were then derived using 20 \fei\ lines from \citet{Placco2015}. No \feii\ lines were detected. 

\textbf{CD$-$38 245} is a red giant star first studied by \citet{bessell1984}. \citet{cayrel2004} determined $T_{\mathrm{eff}}=4800\pm100$\,K following \citeauthor{alonso1999}'s \citeyearpar{alonso1999} color-indices calibration, $\logg=1.5\pm0.1$ from \fei/\feii\ and \ion{Ti}{i}/\ion{Ti}{ii} ionization equilibrium and $\vt=2.2\pm0.1$\,{\kms} by minimizing \fei\ abundance-$EW_{\mathrm{obs}}$ trend. Using these stellar parameters, they deduced $\mbox{[Fe/H](LTE)}=-4.19$.

We determine $\mbox{[Fe/H](LTE)}=-4.28$ and $\mbox{[Fe/H](NLTE)}=-4.03$ using 102 \fei\ and 7 \feii\ lines from \citet{cayrel2004}. Our NLTE stellar parameters are $T_{\mathrm{eff}} =4700$\,K, log\,$g = 2.0$ and $\xi_{\mathrm{t}} = 2.1$\,km\,s$^{1}$. While our \Teff\ agrees with that of \citet{cayrel2004} within error bars, it is worth noting that our \logg\ value is 0.5\,dex higher than theirs. 

\textbf{HE~1310$-$0536 \& HE~2239$-$5019} are two stars also analyzed by \citet{hansen2014}. Their stellar parameters were derived in the same way as that of HE~0233$-$0343. The authors derived $T_{\mathrm{eff}}=5000\pm100$\,K, $\logg=1.9\pm0.3$, $\mbox{[Fe/H](LTE)}=-4.2\pm0.2$ and $\vt=2.2\pm0.3$\,{\kms} for HE~1310$-$0536, and $T_{\mathrm{eff}}=6000\pm100$\,K, $\logg=3.5\pm0.3$, $\mbox{[Fe/H](LTE)}=-4.2\pm0.2$ and $\vt=1.8\pm0.3$\,{\kms} for HE~2239$-$5019 respectively.
 
We determine $\mbox{[Fe/H](LTE)}=-4.25\pm0.18$ and $\mbox{[Fe/H](NLTE)}=-3.77\pm0.12$ from 17 \fei\ lines for HE~1310$-$0536 and $\mbox{[Fe/H](LTE)}=-4.18\pm0.12$ and $\mbox{[Fe/H](NLTE)}=-3.76\pm0.09$ from 15 \fei\ lines for HE~2239$-$5019. Our best fit NLTE stellar parameters agree with those derived by \citet{hansen2014}. The \fei\ NLTE line abundances are found to have no trend with $\chi$. As no \feii\ lines could be detected, we could not test whether the \fei/\feii\ agreement was satisfied. However, future detection and measurements of \feii\ lines can better validate our \logg\ results.

\textbf{SDSS~J0140+2344, CS~30336$-$049 \& HE~2139$-$5432} are three stars analyzed by \citet{norris2013}. The authors used spectro-photometry, Balmer lines fitting and H${\gamma}$ line indices to determine $T_{\mathrm{eff}}=5703\pm60$\,K, $4725\pm60$\,K and $5416\pm41$\,K for SDSS~J0140+2344, CS~30336$-$049 and HE~2139$-$5432, respectively. For SDSS~J0140+2344, no \feii\ lines were measured, and the authors employed a 12 Gyr isochrone to determine two \logg\ values: 4.7 for a dwarf case and 3.4 for a subgiant. Two corresponding values of $0.8$\,{\kms} and $1.5$\,{\kms} for \vt\ and $-4.00$ and $-4.09$ for [Fe/H](LTE) were determined, respectively. For CS~30336$-$049, 3 \feii\ lines were detected, and 1 \feii\ line for HE~2139$-$5432. Using \fei/\feii\ ionization equilibrium, they derived $\logg=1.2$ and $\logg=3.0$ for CS~30336$-$049 and HE~2139$-$5432, respectively. From these parameters, $\vt = 2.1$\,{\kms} and $\mbox{[Fe/H](LTE)}=-4.03$ for CS~30336$-$049 and $\vt = 0.8$\,{\kms} and $\mbox{[Fe/H](LTE)}=-4.03$ for HE~2139$-$5432 were determined. 
 
We employed our fitting parameter technique to determine NLTE stellar parameters for these stars. For SDSS~J0140+2344, we find $T_{\mathrm{eff}}=5600$\,K, $\logg=4.6$ and $\vt=1.0$\,{\kms}. Based upon these results, we favor the dwarf scenario over the subgiant case, since our determined $\logg=4.6$ resulted from a much better fit of the computed $EW$ to the $EW_{\mathrm{obs}}$ than in the subgiant case. Using 35 \fei\ lines from \citet{norris2013}, we subsequently determine $\mbox{[Fe/H](LTE)}=-4.09\pm0.13$ and $\mbox{[Fe/H](NLTE)}=-3.83\pm0.08$ for SDSS~J0140+2344. For CS~30336$-$049 and HE~2139$-$5432, we use 74 \fei\ and 3 \feii\ lines and 32 \fei\ and 1 \feii\ lines from \citet{norris2013} respectively to determine $T_{\mathrm{eff}}=4685$\,K, $\logg=1.4$ and $\vt=2.1$\,{\kms} and $T_{\mathrm{eff}}=5270$\,K, $\logg=3.2$ and $\vt=1.0$\,{\kms}, respectively. It follows that we obtain $\mbox{[Fe/H](LTE)}=-4.22\pm0.21$ and $\mbox{[Fe/H](NLTE)}=-3.91\pm0.16$ for CS~30336$-$049 and $\mbox{[Fe/H](LTE)}=-4.00\pm0.25$ and $\mbox{[Fe/H](NLTE)}=-3.52\pm0.17$ for HE~2139$-$5432.

\textbf{HE~0057$-$5959 \& HE~1424$-$0241} are two evolved red giant stars studied by \citet{cohen2004} and \citet{cohen2008}, respectively. Both studies used color indices to determine $T_{\mathrm{eff}}=5257\pm100$\,K and $T_{\mathrm{eff}}=5195\pm100$\,K for HE~0057$-$5959 and HE~1424$-$0241, and 12 Gyr isochrones to derive $\logg=2.6$ and $\logg=2.5$, respectively. Removing abundances vs. line strength trends, they obtained  $\vt=1.5$\,{\kms} and $\vt=1.8$\,{\kms}. Using these parameters, they derived $\mbox{[Fe/H](LTE)}=-4.08\pm0.2$ and $\mbox{[Fe/H](LTE)}=-4.05\pm0.2$ for HE~0057$-$5959 and HE~1424$-$0241, respectively.

For HE~0057$-$5959, we determine $\mbox{[Fe/H](LTE)}=-4.28\pm0.21$ and $\mbox{[Fe/H](NLTE)}=-3.83\pm0.12$ from 53 \fei\ lines from \citet{cohen2004} (no \feii\ lines were detected). With our fitting method, we determine NLTE $T_{\mathrm{eff}}=5200$\,K, $\logg=2.8$ and $\vt=1.9$\,{\kms}. For HE~1424$-$0241, we use 39 \fei\ and 5 \feii\ lines to determine $\mbox{[Fe/H](LTE)}=-4.19\pm0.20$ and $\mbox{[Fe/H](NLTE)}=-3.73\pm0.15$. The LTE results are in good agreement with \citet{cohen2004} and \citet{cohen2008}. We obtain $T_{\mathrm{eff}} =5140$\,K, $\logg=2.8$ and $\vt=2.2$\,{\kms} for this star.

 \textbf{SDSS J2209$-$0028} is a warm dwarf star \citep{spite2013}. The authors determined $T_{\mathrm{eff}}=6440$\,K using $(g - z)_{0}$ color calibrations from \citet{ludwig2008}. They assumed $\logg=4.0$ which they found to satisfy the \fei/\feii\ ionization equilibrium.  They adopted $\vt=1.3$\,{\kms} and determined $\mbox{[Fe/H](LTE)}=-4.00$.

Using 5 \fei\ lines from \citet{spite2013}, we determine $\mbox{[Fe/H](LTE)}=-3.97\pm0.13$ and $\mbox{[Fe/H](NLTE)}=-3.65\pm0.10$. Due to scarcity of Fe detected lines, we do not determine other NLTE stellar parameters, but instead adopt those from \citet{spite2013}.

\input{fe_final_abundance}

\subsection{Final \fei\ and \feii\ abundances}
We present our final NLTE \fei\ and \feii\ abundances (whenever possible) and their standard deviations ($\sigma_{\mathrm{stdv}}$) in Table~{\ref{Tab:final_abund}}, computed with the spectroscopically determined NLTE stellar parameters given in Table~\ref{tab:ump_list_param}. For comparison, we also present our corresponding LTE values. Despite the scarcity of \fei\ and even more of \feii\ lines, we find that our 4-dimensional $\Teff-\logg-\mbox{[Fe/H]}-\vt$ spectroscopic $EW$ fitting method gives consistent \fei\ and \feii\ abundances to within $0.1$\,dex without having to force this agreement. This adds confidence in our iron atomic model and method used, in addition to the NLTE derived stellar parameter. Additionally, we find slightly smaller standard deviations in the NLTE Fe abundances compared to LTE for most stars. 

Line abundance dispersion can be due to a number of factors including uncertainties in oscillator strengths and other atomic data, $EW$ measurements, model hypotheses (1D/3D, LTE/NLTE, $\dotsc$).  We try in this work to address these possible causes by using the best available gf-values, and include a new approximation of hydrogen inelastic collisions in our NLTE modeling. The scatter is indeed reduced for most stars (See Table~{\ref{Tab:final_abund}}). However, a full 3D, NLTE analysis would likely decrease the scatter even more, but it is still challenging and computationally expensive to employ. In this context, it is encouraging that for SMSS~J0313$-$6708, the only UMP star for which a full 3D, NLTE analysis has been performed, \citet{nordlander2017} report fairly similar 1D NLTE and 3D NLTE results (within 0.2\,dex), whereas much larger differences were obtained between LTE and NLTE models. This highlights that accurate abundances can presently be most efficiently obtained with 1D, NLTE models, such as the present study, whenever reliable atomic data are included. A few other full 3D NLTE calculations should however be performed to confirm this conclusion.

\subsection{Light element enhancement effects on final Fe abundances}\label{alph_enh}
While Fe is usually considered a good proxy of the overall metal content of most stars, UMP stars can have large abundance enhancements in light and $\alpha-$elements such C, N, O, Na, Mg, Si, Ca and Ti relative to iron. These elements can be important electron donors and can thus potentially affect the final Fe derived abundances. The feedback contribution from these elements are customarily treated by using $\alpha$-enhanced input model atmospheres of $\mbox{[$\alpha$/Fe]}=+0.40$ for all stars of $\mbox{[Fe/H]}<-1$.

Some of the most metal-poor stars have [C/Fe] values of 3\,dex or more, with similar O and N abundances, see Table~6 of \citealt{Placco2015} for C and N abundances for this sample of UMP stars. \citet{zhao2016} showed that $\alpha$-elements such as Mg, Ti and Ca maintained constant values of $\sim 0.4$ relative to Fe for $-2.6<\mbox{[Fe/H]}<-1.0$, but with $\mbox{[Ca/Fe]}$ having tendencies to increase below $\mbox{[Fe/H]}=-2.0$, up to 0.6\,dex at $\mbox{[Fe/H]}=-2.6$. Other studies of UMP stars, have shown that this ratio can potentially be higher than the canonical $\mbox{[$\alpha$/Fe]}=+0.40$. Often it also varies from one $\alpha-$element to another. 

A detailed systematic elemental abundance study of UMP stars is therefore needed to quantify how these light element enhancements might affect the iron abundances as well as the stellar parameters of our sample stars. We therefore tested the effects of potential light element enhancements. We thus arbitrarily increased the input metal abundances of the stellar model atmosphere of each UMP star by +0.50\,dex, in addition to the standard $\mbox{[$\alpha$/Fe]}=+0.40$ enhancement.

We then recomputed the final $\mbox{[Fe/H]}$ NLTE abundances, and along the way recorded any potential changes in the spectroscopically determined stellar parameters. 
The changes are found to be \Teff\ independent but \logg\ dependent. Hence, the abundances from \fei\ lines are affected, while those from \feii\ are hardly changing. Cooler stars display stronger effects upon increasing the metal enhancement than warmer stars. For one of the coolest star in our UMP sample, CD$-$38~245 (\Teff=4700\,K), a change of $+0.50$\,dex in the input model metal abundance results in a $0.25$\,dex decrease in the average \fei\ abundances and a slight 0.02\,dex increase in the \feii\ abundances. This corresponds to a compensated $-0.20$\,dex change in \logg.
For a hotter star, HE~2139$-$5432 (\Teff=5270\,K), smaller changes of $-$0.06\,dex were obtained for \fei\ and 0.02\,dex for \feii, while negligible differences were noted for the other stellar parameters, including  \vt\ change of $-0.02$\,\kms. For the even hotter star, J1204+1201 (\Teff=5350\,K), smaller differences of 0.01\,dex for \fei\ were obtained. 

On average, for most UMP stars within $5000<\Teff<7000$\,K, this additional metal abundance enhancement of +0.50\,dex thus causes a slight decrease in the final Fe abundance by at most $\sim 0.05$\,dex. This value is much smaller than the typical error bars and can be thus considered negligible. For cooler stars below 5000\,K, however, the enhancement results in larger decreases in the final \fei\ abundances of typically 0.2-0.3\,dex. If the metal abundance enhancement was pushed to +1.00\,dex, the \fei\ abundance would decrease by 0.3-0.4\,dex.
This decrease in the \fei\ abundances is due to electron pumping in the atmospheric model upon enhancing the metal model metallicity, thus increasing the electron collisional rates by decreasing the photon mean free path, and hereby pushing the abundances lower toward LTE. \feii\ abundances, always being the dominant species, are much less affected by these changes.
We note that the second most iron-poor star HE~1327$-$2326 that is also extremely enhanced in C, N and O relative to Fe, is a warm main-sequence star near the turnoff and thus likely not affected by such model metallicity enhancement changes. The warm giants J1303$-$0019 and HE~0107$-$5240 also with large C, N, and O enhancements, on the other hand, might be affected by this at the 0.3-0.5\,dex level.

\subsection{Uncertainties in stellar parameters}
We report random and systematic uncertainties on \Teff, \logg, [Fe/H] and \vt\ of our stellar parameters in Table~\ref{tab:ump_list_param}.
Our procedure for obtaining stellar parameters results in multiple uncertainties which we discuss in the following:

Our initial stellar parameters are obtained with a non-linear fitting method whose uncertainties arise from the covariance matrix in a \Teff-\logg-[Fe/H]-\vt\ parameter space which reflect how constrained the parameters are by the data (both measured and computed $EW$). Uncertainties depend on the error estimates of the measured $EW$ which typically vary from 1-5\,m{\AA}. Unfortunately, the $EW$ used in this work have been adopted from different reference studies that do not always report $EW$ measurements uncertainties. In those cases, we adopt a nominal value of 2\,m{\AA} for all the lines. This value is typical for low S/N spectra of UMP stars (e.g., \citealp{bonifacio2015,caffau2012}). 
The resulting typical fitting uncertainties for all our stars are  50\,K in \Teff, 0.2\,dex in \logg\ and 0.2\,\kms\ in \vt. We note that employing a higher value for nominal $EW$ measurement uncertainty of 5\,m{\AA} would increase the uncertainties up to 100\,K in \Teff, 0.4\,dex in \logg\ and 0.3\,\kms\ in \vt.
It is important to note that the $1 \sigma$ level fitting uncertainties obtained for \logg\ are underestimated for stars lacking any \feii\ line detections. The \logg\ values obtained from the fitting method for HE~1327$-$2326, HE~0107$-$5240, J1313$-$0019, HE~0233$-$2343, J1204$+$1201, HE~1310$-$0536, HE~2239$-$5019, J0140$+$2344, HE~0057$-$5959 are driven solely by \fei\ lines and thus their uncertainties from the method do not properly reflect the surface gravity dependence on \feii\ lines. We therefore decided to use a constant value of 0.40\,dex as a \logg\ fitting uncertainty for these stars, as an average of the values reported by previous studies for \logg\ of UMP stars with no \feii\ lines (e.g. \citealt{frebel2008,caffau2012,Frebel2015,bonifacio2015}).
Other random uncertainties arise from the uncertainty of the slopes of Fe line abundances versus $\chi$ and $\log(EW/\lambda)$ when obtaining the final effective temperature and \vt\, respectively. Varying the slope within its uncertainty (as determined by the data points) induces a change in the slopes which results in typical uncertainties of 100\,K in \Teff\ and 0.2\,\kms\ in \vt. Individual results ($\sigma^{\mathrm{slope}}$) for each star are listed in Table\,\ref{tab:ump_list_param}.  Random uncertainties of \logg\ are determined from varying the ionization equilibrium of \fei\ and \feii\ within their uncertainties. We adopt the corresponding  change as final \logg\ uncertainty ($\sigma^{\mathrm{var}}$), with typical values of 0.3\,dex. For stars with no \feii\ detection, the same constant value as for the fitting uncertainties of 0.40\,dex was used.

We now report uncertainties in our final \fei\ and \feii\ abundances. First, the dispersion in individual line measurements, quantified by the standard deviation ($\sigma^{\mathrm{stdv}}$).
Typical values are 0.12\,dex in NLTE and 0.20\,dex in LTE for \fei. There are not enough \feii\ lines to meaningfully quantify this for \feii\ so we adopt the \fei\ results instead. We take the standard deviation as our Fe abundance uncertainties because the standard errors of \fei\ would be unrealistically small (e.g., 0.02 and less). ($\sigma^{\mathrm{stdv}}$) are reported in Table\,\ref{Tab:final_abund}.
Second, systematic uncertainties arising from varying the stellar parameters \Teff, \logg\ and \vt\ by about their uncertainty of $\pm 100$\,K, $\pm 0.2$\,cgs and $\pm 0.2$\,\kms\ respectively. The resulting changes in the average Fe abundances typically are $\pm 0.07$\,dex in \fei\ and $\pm 0.01$\,dex for \feii\ for changes in changes in \Teff, $\pm 0.05$\,dex for \fei\ and $\pm 0.2$\,dex for \feii\ for changes in \logg\ and finally $\pm 0.1$\,dex for \fei\ and $\pm 0.02$\,dex for \feii\ for changes in \vt. Total Fe abundance uncertainties are obtained by summing individual uncertainties ($\sigma^{\mathrm{std}}$ and $\sigma^{\mathrm{sys}}$) in quadrature. This leads to a typical total average value of 0.13\,dex.

Similarly, the total uncertainties in the other stellar parameters are obtained by summing individual uncertainties ($\sigma^{\mathrm{fit}}$, $\sigma^{\mathrm{slope}}$, and $\sigma^{\mathrm{var}}$) in quadrature. This leads to typical total uncertainties of 112\,K in \Teff, 0.45\,dex in \logg\ for stars with \feii\ detection and 0.55\,dex for stars without, and 0.4\,\kms in \vt. The individual uncertainties for each star are listed in Table~\ref{tab:ump_list_param}. 
These uncertainties well reflect the challenge of having available only a limited number of Fe lines in these most iron-poor stars.

\section{NLTE corrections}\label{NLTE_corr}
We now discuss the differences between our NLTE and LTE iron abundances [Fe/H] for the UMP stars. We also report the differences between previously determined stellar parameters (\Teff, \logg\ and \vt) from the literature (where either full LTE or partial LTE and photometric methods were used). These NLTE corrections for [Fe/H]  are shown in Table~\ref{Tab:final_abund}, while those for \logg, \Teff\ and \vt\ are listed in Table~\ref{tab:ump_list_param}.

\subsection{\normalfont [Fe/H] abundance corrections}\label{iron_corr}
\begin{figure*}[ht!]
\includegraphics[scale=0.38]{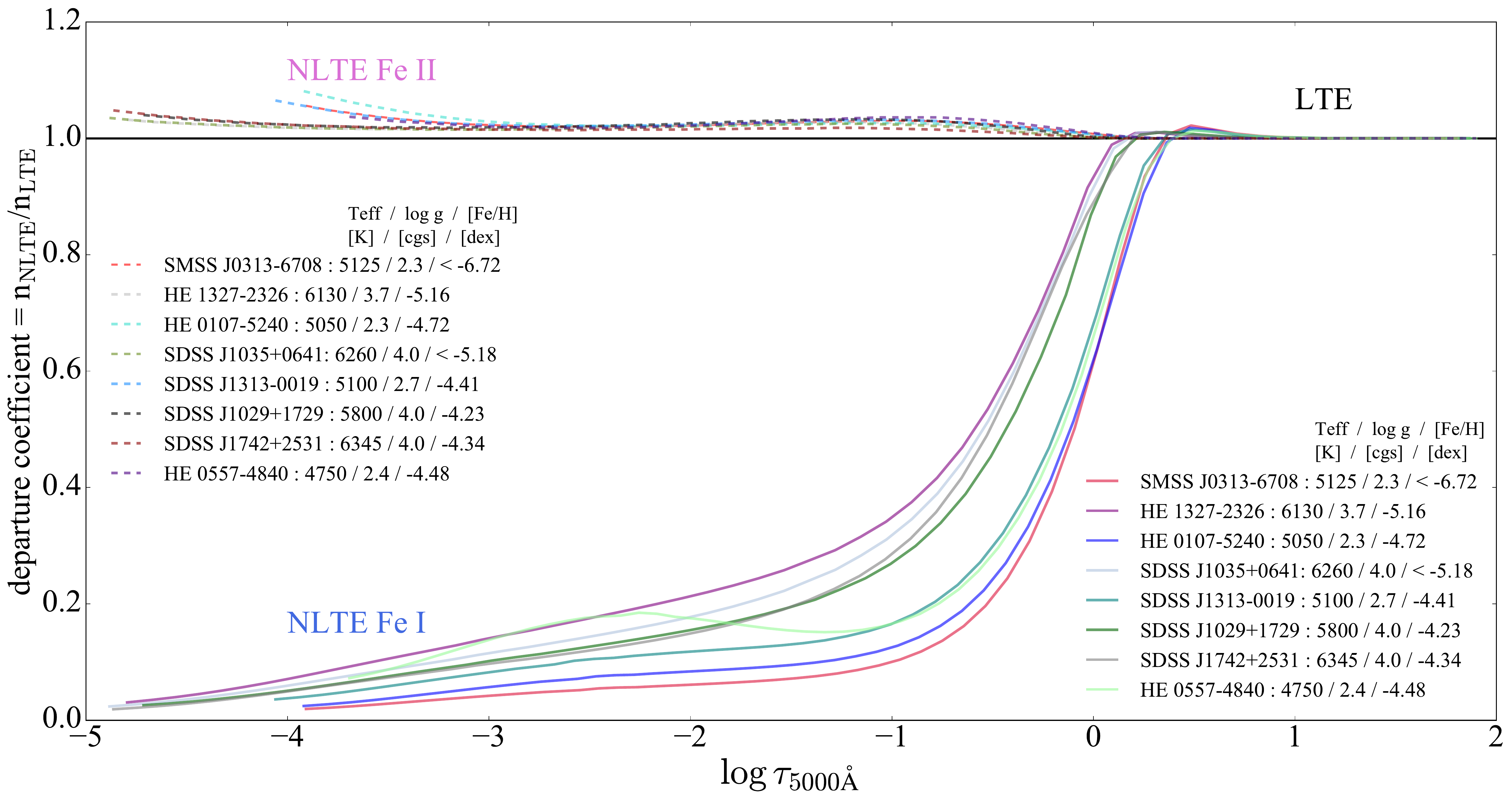}
\caption{\label{dep_coeff}Departure coefficients for eight illustrative UMP stars with $\mbox{[Fe/H]}<-4.00$. Departures are shown for the ground level populations of both \fei\ and \feii\ as a function of optical atmospheric depth $\tau$ at 5000\,{\AA}. \fei\ populations deviate strongly from LTE down to $\mbox{n}_{\mathrm{NLTE}}/\mbox{n}_{\mathrm{LTE}}\sim0.1$ at the lowest [Fe/H] for SMSS~J0313$-$6708, while \feii\ population deviations can be considered negligible compared to \fei.}
\end{figure*}

We define the NLTE Fe line abundance correction for a specific spectral line as the difference between the NLTE and LTE Fe abundance for a given measured equivalent width. 
We calculate $\Delta$[Fe/H] $=$ [Fe/H]$_{\mathrm{NLTE}}$ - [Fe/H]$_{\mathrm{LTE}}$, based on the average abundance differences across all individual Fe lines. The results as well as the number of \fei\ and \feii\ lines used for each UMP star are listed in Table~\ref{Tab:final_abund}. The corrections are found to increase with decreasing [Fe/H] which can be understood due to the increasing magnitude of the over-ionization ($J_{\nu} - B_{\nu}$ excess) in the UV. This over-ionization shifts the ionization-recombination balance towards more efficient ionization, thus de-populating the lower levels relative to LTE. This effect grows larger at lower metallicities as radiative rates become more efficient due to the decrease in electron number densities in the optically transparent atmospheric layers \citep{mashonkina2011,lind2012,mashonkina2016}. The deviation from LTE in the line formation within the depth of the stellar atmosphere can be seen in Figure~\ref{dep_coeff}, where the relative populations (NLTE to LTE) of the ground \fei\ level for the UMP stars with $\mbox{[Fe/H]}<-4.00$ are displayed along their atmospheric depths at 5000\,{\AA} ($\log \, \tau_{5000}$). While the departures from LTE increase with decreasing Fe abundances, other factors such as lower gravities and higher effective temperatures can also play a role in the population deviations from LTE throughout the stellar atmospheres \citep{lind2012,mashonkina2016}.

\begin{figure*}[t!]
\includegraphics[scale=0.35]{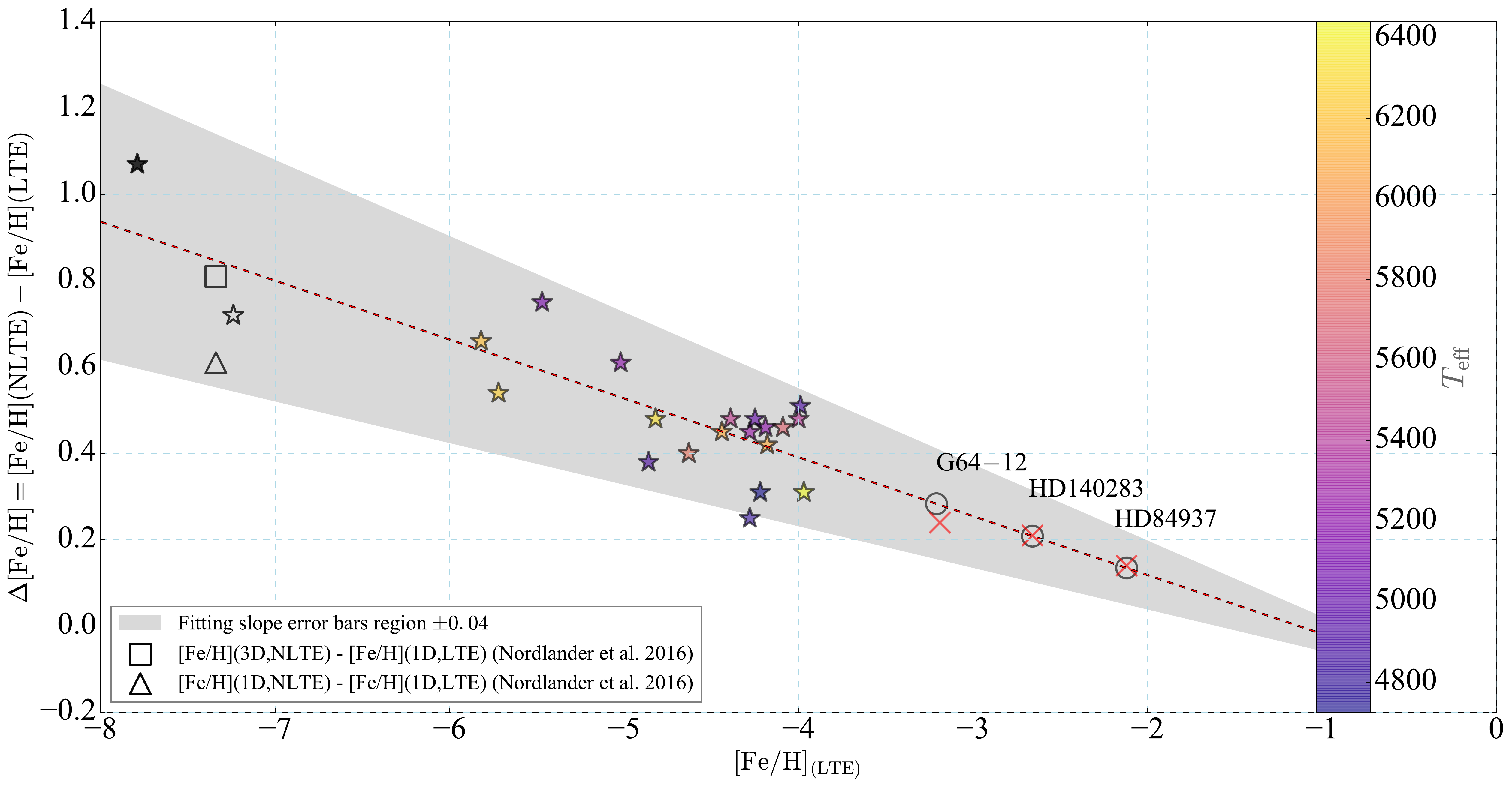}\caption{\label{fig:nlte_corr}Differences between LTE and NLTE Fe abundances as a function of [Fe/H](LTE). UMP stars are shown as star symbols, and an additional three metal-poor reference stars as open circles. \Teff\ [K] for all stars is color-coded. Black stars, the square and the triangle represent different possible NLTE corrections obtained for SMSS J0313$-$6708 as follows: (i) Filled black star is based on using the resonance line at $\lambda$3859.911\,{\AA}. (ii) Open black star is the correction based on the line at $\lambda$3608.859\,{\AA}. (iii) Open square is based on 1D NLTE corrections from \citet{nordlander2017}. (iv) Open triangle is based on the difference between 3D, NLTE to the 1D, LTE results by \citet{nordlander2017} (see text for discussion). The red dashed line represents the fit for the UMP stars excluding SMSS J0313$-$6708.  Open circles correspond to NLTE corrections obtained using Equation~\ref{correction_fit} for the benchmark metal-poor stars HD~140283, HD~84937 and G~64$-$12. Red crosses correspond to the NLTE corrections obtained by \citet{amarsi2016}. The reference stars agree perfectly with the suggested UMP trend of $\Delta$[Fe/H]. All stars lie within the fitting slope error bars region of $\pm0.04$ (gray shaded region).}
\end{figure*}

The NLTE corrections as a function of [Fe/H](LTE) for the UMP stars are shown in Figure~\ref{fig:nlte_corr}. The data are easily fit with a linear relation:

\begin{equation} \label{correction_fit} 
\Delta \mbox{[Fe/H]} = -0.14 \, \mathrm{[Fe/H]}_{\mathrm{LTE}} - 0.15
\end{equation}

The upper limit correction of $\Delta\mbox{[Fe/H]} = 1.07$ for SMSS~J0313$-$6708 was excluded from the fit as no iron lines detection were made in this star. The fitting fitting slope and $\Delta \mbox{[Fe/H]}$-intercept standard errors are respectively $\pm 0.04$ (shown by the gray shaded area in Figure\,\ref{fig:nlte_corr}) and $\pm 0.18$. All the stars, including SMSS~J0313$-$6708, lie within this error bar (gray shaded) region. 

This tight relation allows extending the NLTE corrections to other stars, and potentially also towards higher metallicities ($\mbox{[Fe/H]}>-4.00$). We test this on the benchmark metal-poor stars HD~84937 ($\mbox{[Fe/H](LTE)}=-2.12$), HD~140283 ($\mbox{[Fe/H](LTE)}=-2.66$) and G~64$-$12 ($\mbox{[Fe/H](LTE)}=-3.21$) \citep{amarsi2016}. Using Equation~\ref{correction_fit}, we calculate NLTE corrections of 0.14\,dex, 0.22\,dex and 0.29\,dex for HD~84937, HD~140283 and G~64$-$12, respectively. \citet{amarsi2016} studied these three stars using a full 3D and 1D NLTE analyses, using for the first time quantum mechanical atomic data for hydrogen collisions, and reliable non-spectroscopic atmospheric parameters. The authors report 0.14\,dex and 0.21\,dex and 0.24\,dex as 1D NLTE corrections for HD~84937, HD~140283 and G~64$-$12, respectively. These values are in excellent agreement with our values. Our fit can thus be used to predict NLTE corrections of metal-poor stars though the whole range of metallicities [Fe/H] from at least -8.00 to -2.00\,dex, which further asserts that our relation can be used and applied to LTE Fe abundances of a variety of metal-poor stars.

\subsection{Consequences for spectroscopic determination of stellar parameters \Teff\ and \logg}
We present in Table~{\ref{Tab:final_abund}} the difference in stellar parameters \Teff, \logg\ and \vt\
between our NLTE and previously derived LTE spectroscopic or photometric values, whenever possible. This illustrates the changes by going to a full NLTE Fe line analysis. 
We obtain positive $\Delta \logg = \logg\ \mbox{(NLTE)} - \logg\ \mbox{(lit. value)}$ of 0.1 - 0.5\,dex for all UMP stars whenever a NLTE \logg\ derivation was possible. An important consequence is that surface gravities derived by LTE analyses tend to be lower than what is expected in NLTE. LTE values should thus be corrected before any further elemental abundance determination.
Our positive NLTE \logg\ corrections are in agreement with previous studies, e.g., \citet{thevenin1999} who have found positive $\Delta \logg$ for a large number of metal-poor stars. Their values were found to be in agreement with spectroscopic independent \logg\ determinations, e.g. those derived from HIPPARCOS parallaxes. 

Our NLTE \Teff\ agree within error bars with the those photometrically derived values using color-\Teff\ calibration relations (e.g. \citealt{alonso1999,alonso2001}).
However, comparing values spectroscopically determined through LTE \fei\ excitation equilibrium with our results shows deviations of up to $\sim 150$\,K, where NLTE \Teff\ values are lower ($\Delta \Teff = \Teff\ \mbox{(NLTE)} - \Teff\ \mbox{(lit. value)} < 0$). These deviations can be expected as the LTE Boltzmann equilibrium of atoms cannot pertain at lower metallicities, especially for the non-dominant neutral Fe species, which will affect the excitation balance of \fei\ lines.

We treat the microturbulent velocity \vt\ as a free parameter but  do not further consider the obtained results for $\Delta \vt = \vt\ \mbox{(NLTE)} - \vt\ \mbox{(lit. value)}$. Nevertheless, we report the obtained values in Table.~\ref{Tab:final_abund}.

Given that there is a strong correlation of  $\Delta\mbox{[Fe/H](LTE)}$ as a function of [Fe/H], we also attempted to determine $\Delta\mbox{[Fe/H]}$ as a function of \Teff\ and \logg. This requires inspecting lines of similar strengths ($EW$), and lines with the same lower levels excitation potential $\chi$, as abundances derived from similar lines depend on the thermal stratification of the atmosphere. For example, low excitation lines are mostly prone to 3D effects which can lower \fei\ abundances of metal-poor stars by $0.1$\,dex \citep{amarsi2016}. 
However, due to the small number of UMP stars, and scarcity of Fe lines, there is not enough data available to map out temperature and surface gravity dependent $\Delta\mbox{[Fe/H]}$.
It is thus rather difficult to quantify the dependence of NLTE effects on other parameters than [Fe/H] in our sample of stars. If more stars are found in ongoing and future surveys, this question should be revisited.
Nevertheless, spectroscopically derived stellar parameters using the LTE formalism have to be corrected for NLTE effects.

\newpage
\section{Conclusions}\label{conc}
We have presented 1D, NLTE Fe line-by-line formation computations for 20 UMP stars with $\mbox{[Fe/H](LTE)}\lesssim-4.0$. We use NLTE \fei\ and \feii\ lines abundances, when available, to also determine spectroscopic stellar parameters \Teff, \logg\ and \vt\, in addition to [Fe/H]. Our results show that:

\begin{itemize}
\item[-] Our NLTE Fe abundance corrections for the UMP stars are larger than any previous determinations, up to $\sim 1.00$\,dex at the lowest iron abundances. These results set a new scale of NLTE corrections to be applied to LTE abundances of other metal-poor stars. The larger corrections are mainly due to performing a full NLTE analysis, using new estimates of hydrogen collision rates and the inclusion of charge transfer rates for the first time in the NLTE analysis of UMP stars. 

\item[-]The line-by-line abundance scatter in NLTE is decreased for most stars down to $\sigma^{\mathrm{stdv}} \sim 0.12$\,dex as compared to LTE.

\item[-]The NLTE corrections we calculated over the $-7.00<\mbox{[Fe/H](LTE)}<-4.00$ range can be extrapolated up to at least $\mbox{[Fe/H]}=-2.00$, to predict NLTE corrections, in perfect agreement with independent (1D and 3D, NLTE) determinations.

\end{itemize}

Even though  the number of known UMP stars  has greatly increased over the last few years to a sample of 20 stars, the relatively small number remains a shortcoming to a full stellar population analysis. Future surveys are expected to deliver additional UMP stars, hopefully extending to $\mbox{[Fe/H]}\sim-7.0$.
However, now is the time to revisit existing data and to analyze the known stars as precisely and uniformly as possible, as is presented for Fe abundances in this work.

Our results provide a fist step towards a full NLTE chemical species analysis of UMP and EMP stars. A full NLTE abundance pattern will enable us to put constraints on the Initial Mass Function (IMF) and other properties of Pop III stars, by comparing accurately computed NLTE abundances of a full set of elements to model supernova yields (e.g., as has been done in LTE by \citealt{Placco2015}).

\acknowledgements
We thank Piercarlo Bonifacio for the useful conversation and advice. We are thankful to Terese Hansen for providing us with equivalent width measurements for some of our studied stars. R.E. acknowledges support from a JINA-CEE fellowship (Joint Institute for Nuclear Astrophysics - Center for the Evolution of the Elements), funded in part by the National Science Foundation under Grant No. PHY-1430152 (JINA-CEE). A.F. is supported by NSF-CAREER grant AST-1255160. AF acknowledges support from the Silverman (1968) Family Career Development Professorship. 
This work was supported in part by the French-Lebanese Program PHC Cedre. This work has made use of the VALD database, operated at Uppsala University, the Institute of Astronomy RAS in Moscow, and the University of Vienna.

\bibliography{ref}


\clearpage
\include{ump_linelists_stub}
\end{document}

%% file: Stellar_params.tex
\begin{deluxetable*}{c l c c c c c c c c c c c c c}
\tablewidth{0pt}
\tabletypesize{\footnotesize}
\tablecaption{\label{tab:ump_list_param} Derived stellar parameters of ultra metal-poor stars with their slope ($\sigma^{\mathrm{slope}}$, for \Teff\ and \vt), variation ($\sigma^{\mathrm{var}}$, for \logg) and fitting ($\sigma^{\mathrm{fit}}$, for \Teff, \logg\ and \vt) uncertainties on their values (see Section\,4.6 for detailed description of each uncertainty). For stars with not enough Fe lines to derive spectroscopic atmospheric parameters, corresponding literature values were adopted (indicated by the table notes below). Columns 11-13 present the NLTE corrections obtained from the differences between our spectroscopic NLTE stellar parameters and previously derived LTE or photometric values. Column 14 shows the NLTE Fe abundance corrections derived in this work.}
\tablehead{
\colhead{Star} & \colhead{\Teff} & \colhead{$\sigma^{\mathrm{slope}}$}  & \colhead{$\sigma^{\mathrm{fit}}$} & \colhead{\logg} & \colhead{$\sigma^{\mathrm{var}}$}  & \colhead{$\sigma^{\mathrm{fit}}$} & \colhead{\vt} & \colhead{$\sigma^{\mathrm{slope}}$}  & \colhead{$\sigma^{\mathrm{fit}}$} & & \colhead{$\Delta$\Teff} & \colhead{$\Delta$\logg}  & \colhead{$\Delta$\vt} & \colhead{$\Delta$[Fe/H]}
\\
& \colhead{[K]} & \colhead{[K]}  & \colhead{[K]}  & \colhead{[cgs]} & \colhead{[cgs]} & \colhead{[cgs]} & \colhead{[\kms]} & \colhead{[\kms]} & \colhead{[\kms]} & & \colhead{[K]} & \colhead{[cgs]}  & \colhead{[\kms]} & \colhead{[dex]}}
\startdata
\multicolumn{14}{c}{$-4.5 < \mbox{[Fe/H]} <-4.0$} \\
\hline
SDSS J2209$-$0028 & 6440\tablenotemark{1} & \nodata & \nodata & { }4.0\tablenotemark{2} & \nodata & \nodata &  1.3 & \nodata & \nodata & &  \nodata    & \nodata     & \nodata &0.32\\
HE 2139$-$5432    & 5270        & 100     & 43      & 3.2     & 0.30    & 0.15    &  1.0     & 0.2     & 0.2     &  &  $-146$   & $0.20$     &$0.2$                  & 0.48\\
CS 30336$-$049    & 4685        & 80      & 35      & 1.4     & 0.30    & 0.22    &  2.1     & 0.2     & 0.1     &  & $-40$    & 0.20        &\nodata                & 0.31\\
HE 1424$-$0241    & 5140        & 60      & 46      & 2.8     & 0.40    & 0.37    &  2.2     & 0.3     & 0.2     &  & $-55$  & 0.30        & $0.4$                 & 0.46\\
HE 0057$-$5959    & 5200        & 110     & 68      & 2.8     & 0.40    & 0.40    &  1.9     & 0.2     & 0.3     &  & $-57$    & $0.20$      & 0.4                   & 0.45 \\
SDSS J0140+2344   & 5600        & 100     & 77      & 4.6     & 0.40    & 0.40    &  1.0     & 0.5     & 0.2     &  & $-103$   & \nodata     & \nodata & 0.26 \\
HE 2239$-$5019    & 6000        & 80      & 49      & 3.5     & 0.40    & 0.40    &  1.8     & 0.2     & 0.1     & &  \nodata   & \nodata     &\nodata                & 0.42  \\
HE 1310$-$0536    & 5000        & 70      & 43      & 1.9    & 0.40    & 0.40    &  2.2     & 2.0     & 1.0     &  & \nodata  & \nodata      & \nodata               & 0.48  \\
 CD$-$38 245      & 4700        & 60      & 38      & 2.0    & 0.40    & 0.21    &  2.1     & 0.2     & 0.1     &  & $-100$   & 0.50         & \hspace{-0.23cm}$-0.1$& 0.25  \\
 SDSS J1204+1201  & 5350        & 100     & 45      & 3.3     & 0.40    & 0.40    &  1.5     & 0.2     & 0.2     &  & $-117$   & $0.10$      &\nodata                & 0.48  \\
 CS 22949$-$037   & 4800        & 90      & 67      & 1.9     & 0.30    & 0.20    &  1.9     & 0.2     & 0.2     &  & $-100$   & 0.40        & \hspace{-0.23cm}$0.1$& 0.51 \\
\hline
\multicolumn{14}{c}{$-5.0 < \mbox{[Fe/H]}< -4.5$} \\
\hline
 HE 0233$-$0343  & 6020  & 80      & 52      & 3.4    & 0.40    & 0.40    &  1.8    & 0.3     & 0.2     &  & $-80$   & \nodata  & \nodata & 0.45\\
 HE 0557$-$4840  & 4800  & 80      & 67      & 2.4    & 0.30    & 0.49    &  1.8    & 0.4     & 0.3     &  & $-100$  & 0.20     &\nodata & 0.38\\
 SDSS J1742+2531 & 6345\tablenotemark{3} & \nodata & \nodata & { }4.0\tablenotemark{4} & \nodata & \nodata &  1.5  & \nodata & \nodata & &  \nodata &  \nodata  & \nodata & 0.48\\
 SDSS J1029+1729 & 5811\tablenotemark{5} & \nodata & \nodata & { }4.0\tablenotemark{6} & \nodata & \nodata &  1.5  & \nodata & \nodata &  &  \nodata & \nodata  &\nodata & 0.40\\
\hline
\multicolumn{14}{c}{$\mbox{[Fe/H]}< -5.0$} \\
\hline
 SDSS J1313$-$0019 & 5100  & 80      & 67        & 2.7  & 0.40    & 0.40   &  1.8   & 0.2     & 0.2  &   &   $-70$   & 0.10      & \nodata& 0.61\\
 SDSS J1035$+$0641 & 6260\tablenotemark{3} & \nodata & \nodata   & { }4.0/4.4\tablenotemark{4} & \nodata & \nodata & 1.5  & \nodata & \nodata & & \nodata & \nodata   & \nodata & 0.54\\  
 HE 0107$-$5240    & 5050  & 60      & 43        & 2.3   & 0.40    & 0.40    &  2.2   & 0.3     & 0.3   &  &  $-50$   & 0.10   & \nodata & 0.75\\
 HE 1327$-$2326  & 6130                  & 100      & 32        & 3.7                  & 0.40    & 0.40    &  1.7    & 0.4     & 0.3     & &  $-50$   & \nodata   & 0.4 & 0.66\\
 SMSS J0313$-$6708 & 5125\tablenotemark{7} &\nodata  & \nodata   & { }2.3\tablenotemark{7} & \nodata & \nodata &  2.0  &\nodata  & \nodata & &
 \nodata &  \nodata   & \nodata & $\mathbf{1.07}$\\
\enddata
\tablenotetext{1}{From photometry; \citet{spite2013}}
\tablenotetext{2}{Fixed adopted value; \citet{spite2013}}
\tablenotetext{3}{From photometry and H-$\alpha$ wings fitting; \citet{bonifacio2015}}
\tablenotetext{4}{From 12 Gyr isochrone; \citet{bonifacio2015}}
\tablenotetext{5}{From photometry; \citet{caffau2012}}
\tablenotetext{6}{From \ion{Ca}{i}/\ion{Ca}{ii} ionization equilibrium; \citet{caffau2012}}
\tablenotetext{7}{From spectrophotometry and H line profiles fitting; \citet{bessell2015}}
\end{deluxetable*}

%% file: fe_final_abundance.tex
\begin{deluxetable*}{c c c c c c  c c c}
\tablewidth{0pt}
\tabletypesize{\footnotesize}
\tablecaption{\label{Tab:final_abund} Our results for the average \fei, \feii\ and total Fe abundances with their standard deviation errors obtained for the UMP stars both in LTE and NLTE respectively. The number of \fei\ and \feii\ lines used in the analysis are shown in the last two columns. Values are reported relative to the solar iron abundance of $\epsilon(\mathrm{Fe})_{\odot}=7.50$ from \citet{asplund2009}.}
\tablehead{
\colhead{Star} & \colhead{[\fei/H]$^{\mathrm{LTE}}$} & \colhead{[\feii/H]$^{\mathrm{LTE}}$}  & \colhead{[\fei/H]$^{\mathrm{NLTE}}$} & \colhead{[\feii/H]$^{\mathrm{NLTE}}$} & \colhead{[Fe/H]$^{\mathrm{LTE}}$} & \colhead{[Fe/H]$^{\mathrm{NLTE}}$}  & \colhead{N \fei} & \colhead{N \feii}}  
\startdata
 \multicolumn{9}{c}{$\mathbf{-4.5<\mbox{[Fe/H]}<-4.0}$} \\
\hline
SDSS J2209$-$0028 & $-3.97\pm0.13$ & \nodata        & $-3.65\pm0.10$ & \nodata        &   $-3.97 \pm 0.13$ &  $-3.65 \pm 0.10$  & 5  & 0 \\
HE 2139$-$5432    & $-4.01\pm0.25$ & $-3.53$        & $-3.52\pm0.17$ & $-3.54$        &   $-4.00 \pm 0.25$ &  $-3.52 \pm 0.17$  & 32 & 1 \\
CS 30336$-$049    & $-4.23\pm0.21$ & $-3.83\pm0.08$ & $-3.91\pm0.16$ & $-3.86\pm0.07$ &   $-4.21 \pm 0.20$ &  $-3.91 \pm 0.16$  & 74 & 3 \\
HE 1424$-$0241    & $-4.21\pm0.20$ & $-3.72\pm0.20$ & $-3.74\pm0.11$ & $-3.71\pm0.15$ &   $-4.19 \pm 0.20$ &  $-3.73 \pm 0.15$   &  39 & 5 \\
HE 0057$-$5959    & $-4.28\pm0.21$ & \nodata        & $-3.83\pm0.12$ & \nodata        &   $-4.28 \pm 0.21$ &  $-3.83 \pm 0.12$  & 53 & 0 \\
SDSS J0140+2344   & $-4.09\pm0.13$ & \nodata & $-3.83\pm0.08$ & \nodata &   $-4.09 \pm 0.13$ &  $-3.83 \pm 0.08$  & 35 & 0 \\
HE 2239$-$5019    & $-4.18\pm0.12$ & \nodata        & $-3.76\pm0.09$ & \nodata        &   $-4.18 \pm 0.12$ &  $-3.76 \pm 0.09$  & 15 & 0 \\
HE 1310$-$0536    & $-4.25\pm0.18$ & \nodata        & $-3.77\pm0.12$ & \nodata        &   $-4.25 \pm 0.18$ &  $-3.77 \pm 0.12$  & 17 & 0 \\
 CD$-$38 245      & $-4.28\pm0.20$ & $-4.16\pm0.12$ & $-4.03\pm0.14$ & $-4.09\pm0.12$ &   $-4.28 \pm 0.20$ &  $-4.03 \pm 0.12$  & 102 & 7 \\
 SDSS J1204+1201  & $-4.39\pm0.12$ & \nodata        & $-3.91\pm0.11$ & \nodata        &   $-4.39 \pm 0.12$ &  $-3.91 \pm 0.11$  & 20 & 0 \\
 CS 22949$-$037   & $-3.99\pm0.15$ & $-3.56\pm0.10$ & $-3.44\pm0.12$ & $-3.50\pm0.09$ &   $-3.99 \pm 0.16$ &  $-3.48 \pm 0.13$ & 65 & 5 \\
 \hline
 \multicolumn{9}{c}{$\mathbf{-5.0<\mbox{[Fe/H]}<-4.5}$} \\
 \hline
 HE 0233$-$0343   & $-4.44\pm0.08$ &\nodata  & $-3.99\pm0.08$ & \nodata  &    $-4.44 \pm 0.08 $ & $-3.99 \pm 0.08$ &  11 & 0 \\   
 HE 0557$-$4840   & $-4.86\pm0.17$ & $-4.70$ &$-4.48\pm0.13$  & $-4.52$  &    $-4.86 \pm 0.17 $ & $-4.48 \pm 0.13$ & 59 & 1 \\   
 SDSS J1742+2531  & $-4.82\pm0.07$ & \nodata & $-4.34\pm0.03$ & \nodata  &    $-4.82 \pm 0.07 $ & $-4.34 \pm 0.03$ &  3  & 0 \\   
 SDSS J1029+1729  & $-4.63\pm0.13$ & \nodata & $-4.23\pm0.14$ & \nodata  &  $-4.63 \pm 0.13 $ & $-4.23 \pm 0.14$ &   3  & 0 \\   
 \hline
 \multicolumn{9}{c}{$\mathbf{\mbox{[Fe/H]}<-5.0}$} \\
 \hline
 SDSS J1313$-$0019 & $-5.02\pm0.10$     & \nodata  & $-4.41\pm0.09$     & \nodata         & $-5.02  \pm 0.10$ & $-4.41\pm0.09$ & 36  & 0  \\
 SDSS J1035$+$0641 & $<-5.72\pm 1\sigma$     & \nodata          & $<-5.18\pm1\sigma$     & \nodata         & $<-5.72  \pm 1\sigma  $ &  $<-5.18 \pm 1\sigma$    &   { }1$^{a}$   & 0  \\
 HE 0107$-$5240    & $-5.47\pm0.20$     & $<-4.70$  & $-4.72\pm0.15$     & $<-4.71$ & $-5.47\pm0.20$ &  $-4.72\pm0.15$ & 25 & { }1$^{a}$  \\
 HE 1327$-$2326    & $-5.82\pm0.16$     & $<-5.11$         & $-5.16\pm0.12$     & $<-5.10$        & $-5.82  \pm 0.16$ &  $-5.16 \pm 0.12$         &    10  & { }1$^{a}$  \\
 SMSS J0313$-$6708 & $<-7.79\pm3\sigma$ & \nodata          & $<-6.72\pm3\sigma$ &\nodata          & $<-7.79 \pm 3\sigma$ &  $<-6.72  \pm 3\sigma$ &   { }{} 1$^{a}$   & 0  \\
\enddata
\tablenotetext{a}{No line detection, upper limit only.}
\end{deluxetable*}

%% file: ump_linelists_stub.tex
\begin{deluxetable*}{l l c c r c c c}
\tablewidth{0pt}
\tabletypesize{\scriptsize}
\tablecaption{\label{Tab:abund} Atomic linelists and obtained \fei\ and \feii\ abundances of the UMP stars.}
\tablehead{                                                                                                                                          
\colhead{Star name}& \colhead{Ion} & \colhead{$\lambda$ }&\colhead{$\chi$}&\colhead{log\,$gf$}&\colhead{$EW$}&\colhead{$\log \varepsilon$(X)$_{\mathrm{LTE}}$}&\colhead{$\log \varepsilon$(X)$_{\mathrm{NLTE}}$}  \\
\colhead{}&\colhead{}&\colhead{[\AA] }&\colhead{[eV]}&\colhead{ }&\colhead{[m\AA]}&\colhead{[dex]}&\colhead{[dex]}}
\startdata
SDSS J2209$-$0028 & \fei &   4045.810  &1.48  &0.28   &36.4  &3.50&  3.82\\
&\fei &   4063.590  &1.56  &0.07   &31.7  &3.67 &3.95    \\
&\fei &   4071.740  &1.61  &$-$0.02   &28.9  &3.72 &3.99  \\
&\fei &   4383.549  &1.48  &0.20   &28.8  &3.36 &3.72      \\
&\fei &   4404.750  &1.56  &$-$0.14   &17.4  &3.46 &3.78  \\  
\enddata
\tablecomments{This table is published in its entirety in the machine-readable format.
      A portion is shown here for guidance regarding its form and content.}
\end{deluxetable*}